\documentclass{article}

    \PassOptionsToPackage{numbers,compress}{natbib}

\usepackage[preprint]{neurips_2024}

\usepackage[utf8]{inputenc} 
\usepackage[T1]{fontenc}    
\usepackage[pagebackref=true,breaklinks=true,letterpaper=true,colorlinks,bookmarks=false]{hyperref}
\usepackage{url}            
\usepackage{booktabs}       
\usepackage{amsfonts}       
\usepackage{nicefrac}       
\usepackage{microtype}      
\usepackage{xcolor}         
\usepackage{amsmath,amssymb}
\usepackage{graphicx}
\usepackage{wrapfig}
\usepackage{algorithm,algorithmic}
\usepackage{multirow}
\usepackage{colortbl}
\usepackage{gensymb}
\usepackage{pifont}       
\usepackage{bbding}       
\usepackage{fontawesome}

\usepackage{subcaption}

\title{Blaze3DM: Marry Triplane Representation with Diffusion for 3D Medical Inverse Problem Solving}

\author{%
  Jia He \\
  State Key Laboratory of Multimodal Artificial\\
  Intelligence Systems\\
  Institute of Automation\\
  Chinese Academy of Sciences\\
  \texttt{hejia2020@ia.ac.cn}
  \And
  Bonan Li \\
  School of Electronic, Electrical \\
  and Communication Engineering\\
  University of Chinese Academy of Sciences\\
  \texttt{libonan@ucas.ac.cn} \\
  \AND
  Ge Yang \\
  State Key Laboratory of Multimodal Artificial\\
  Intelligence Systems\\
  Institute of Automation\\
  Chinese Academy of Sciences\\
  \texttt{ge.yang@ia.ac.cn} \\
  \And
  Ziwen Liu\thanks{Corresponding Author} \\
  School of Artificial Intelligence\\
  University of Chinese Academy of Sciences \\
  \texttt{liuziwen@ucas.ac.cn} \\
}

\begin{document}
\maketitle

\begin{figure}[h!]
    \centering
    \includegraphics[width=1\linewidth]{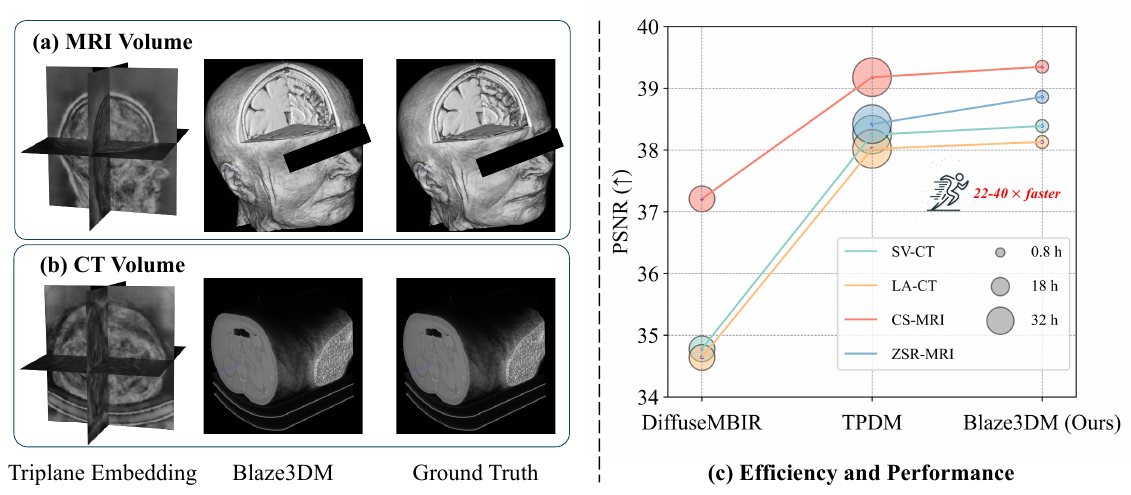}

    \caption{(a) and (b): Volume generation by triplane-based diffusion model of our Blaze3DM on MRI and CT; (c) Blaze3DM speeds up inference time by up to $22\times$ compared to DiffuseMBIR~\cite{chung2023solving} and TPDM~\cite{lee2023improving} while achieving state-of-the-art performance in four tasks.} 
    \label{fig:0}
\end{figure}

\begin{abstract}
    Solving 3D medical inverse problems such as image restoration and reconstruction is crucial in modern medical field. However, the curse of dimensionality in 3D medical data leads mainstream volume-wise methods to suffer from high resource consumption and challenges models to successfully capture the natural distribution, resulting in inevitable volume inconsistency and artifacts. Some recent works attempt to simplify generation in the latent space but lack the capability to efficiently model intricate image details. 
    To address these limitations, we present \textbf{Blaze3DM}, a novel approach that enables fast and high-fidelity generation by integrating compact triplane neural field and powerful diffusion model. In technique, Blaze3DM begins by optimizing data-dependent triplane embeddings and a shared decoder simultaneously, reconstructing each triplane back to the corresponding 3D volume. To further enhance 3D consistency, we introduce a lightweight 3D aware module to model the correlation of three vertical planes. Then, diffusion model is trained on latent triplane embeddings and achieves both unconditional and conditional triplane generation, which is finally decoded to arbitrary size volume. 
    Extensive experiments on zero-shot 3D medical inverse problem solving, including sparse-view CT, limited-angle CT, compressed-sensing MRI, and MRI isotropic super-resolution, demonstrate that Blaze3DM not only achieves state-of-the-art performance but also markedly improves computational efficiency over existing methods (22\textasciitilde40$\times$ faster than previous work). 
\end{abstract}

\section{Introduction}

Medical image inverse problem, the task of reconstructing complete image information from limited observed measurements, is a long-standing problem in the medical computer vision community owing to its ill-posed property, such as sparse-view computed tomography (SV-CT)~\cite{chung2022improving,song2021solving}, limited-angle computed tomography (LA-CT)\cite{chung2022improving,song2021solving}, compressed-sensing magnetic resonance imaging (CS-MRI)~\cite{chung2022come,chung2022score}, etc. When it rises to 3D, the types of inverse problems have been slightly expanded, for example, the resolution degradation in the slice dimension leads to MRI isotropic reconstruction (ZSR-MRI) task\cite{chung2022improving}. Moreover, the difficulty of problem lies in the high dimensionality and complicated distribution of 3D volume data, which presents challenges to high image reconstruction quality and high computational efficiency. This underscores the crucial need for an expressive and efficient representation beyond volume values. It not only promotes the research and application of 3D medical image inverse problem solving but also has broad significance in 3D medical image processing field.

Initial efforts focus on slice-wise methods~\cite{nie2018medical,yu2019ea}, which involve generating and stacking individual slices into a 3D volume using various backbone models. However, these methods often fail to effectively capture the 3D consistency of biomedical structures, leading to significant content jitter along the slice dimension. In response to this limitation, more recent works~\cite{chung2023solving,peng2023generating} have incorporated inter-slice constraints to model dependencies along the $z$-axis, building upon the slice-wise approach. Nevertheless, these methods remain suboptimal as they struggle to learn long-range dependencies across all dimensions, ultimately hindering their ability to accurately model the complete 3D joint distribution. Benefiting from the recent development of generative models, subsequent researches~\cite{lee2023improving,dorjsembe2022three,kim2022diffusion,yu20183d,cirillo2021vox2vox} delve into volume-wise models, directly constructing 3D generative backbones, including Generative Adversarial Networks (GANs)~\cite{gan} and Diffusion Models (DMs)~\cite{ho2020denoising}. However, the substantial computational costs associated with volumetric data have constrained their further development.
Alternatively, utilizing the compact latent feature space seems promising for reducing computational complexity~\cite{zhu2023make,khader2023denoising,pinaya2022brain,rombach2022high,kim2024adaptive}, yet it heavily depends on high-quality feature encoders pre-trained with massive data, which are not always readily available.

In this study, we alleviated the limitations of previous works and proposed a new way to simultaneously achieve high quality and fast speed. The key idea is to construct a compact representation space and perform controllable generation within it. Compared to the ordinary latent space, triplane representation not only can be rapidly and stably trained via a lightweight decoder but also offers more compact and expressive capabilities for spatial position rendering,  which is highly suitable for modeling 3D medical volumes. For the generation part, we introduce the diffusion model that effectively captures complex distributions as the foundational model. Building upon it, our method yields high-quality synthesis results in a highly efficient way.

Particularly, our approach, termed Blaze3DM, operates in two steps. Firstly, we jointly optimized a shared decoder and data-dependent triplane embeddings. The triplanes consist of three kinds of feature maps to represent the information on XY/YZ/XZ planes respectively. To enhance 3D consistency between each plane, a lightweight yet effective 3D aware module is introduced into decoder. Secondly, learned triplane embeddings are employed as inputs to optimize a condition-free diffusion model. During the sampling phase, the diffusion model can be guided by degradation constraints to generate complete triplane representation, which is ultimately mapped to arbitrary size 3D volume using previously mentioned decoder. 

Fig.\ref{fig:0}(a) and (b) illustrate the triplane representation and the generated medical volume using our Blaze3DM method. Fig.\ref{fig:0}(c) compares the efficiency and performance across four tasks, demonstrating that Blaze3DM not only achieves state-of-the-art performance but also significantly reduces the inference time from 10-20 hours to within 1 hour.


In conclusion, our proposed Blaze3DM has threefold core contributions:
\begin{itemize}

    \item To the best of our knowledge, Blaze3DM is the first work to utilize a triplane neural field to model 3D medical volume distributions. The triplane provides an efficient representation by compacting 3D information into the XY, YZ, and XZ planes, while the neural field enables good scalability to different volume sizes based on its continuous function.
    
    \item Building upon the vanilla triplane neural field, Blaze3DM introduces a 3D-aware module that captures the correlations between triplanes, effectively enhancing expressiveness and maintaining volume consistency. Additionally, we employ diffusion models as the generative backbone to achieve high-quality and controllable generation. In particular, we introduce the 2D degradation guidance-based sampling method to solve 3D medical inverse problem.
    
    \item Experiments on four classic 3D medical inverse problem tasks, including SV-CT, LA-CT, CS-MRI, and ZSR-MRI, consistently demonstrate that Blaze3DM achieves state-of-the-art performance while remarkably improving model inference efficiency, accelerating by up to 22\textasciitilde40 times compared to previous work.

\end{itemize}

\section{Relative Work}

\subsection{Triplane Neural Field}

Neural fields represent the 3D objects or volumes as continuous functions, making them easy to edit and scale well with volume complexity. However, original implicit methods require a single network to build the neural field of a single object\cite{mildenhall2021nerf} or volume\cite{wu2021irem}, resulting in inefficiency. Recently, hybrid explicit-implicit representation methods~\cite{chan2022efficient,boulch2022poco,genova2020local,genova2019learning,martel2021acorn} have been widely developed, which draw support from explicit representation and enable small, efficient MLPs to inference, demonstrating better performance in capture details. Among them, triplane representation takes advantage of its efficiency and flexible to integrate with expressive generator architectures. To achieve zero-shot triplane generation, \citet{chan2022efficient} conducted a 3D Generative Adversarial Network (GAN) to generate the triplane representation. Following the popularity of diffusion modeling, \citet{shue20233d,wang2023rodin} advanced the 2D diffusion model as the triplane generation backbone. However, due to significant information compression, simple triplane neural field is still lagging behind in 3D synthesis quality. To improve it, \citet{khatib2024trinerflet} explored and proposed a 2D wavelet-based multiscale triplane representation to improve the generation quality. Additionally, \citet{wu2023sin3dm} introduced Sin3DM, which introduces a convolution-based encoder to extract more information from raw data and utilize 2D convolution block in decoder to improve the reconstruction quality. 

\subsection{Diffusion Models}

\label{sec2-1}
Recently, Diffusion Models (DMs)~\cite{ho2020denoising,song2020denoising,dhariwal2021diffusion,nichol2021improved} have shown excellent performance in image generation. DMs define the \textit{forward process} as a Markov chain that adds random noise with a given variance schedule $\{\beta_t\}_{t=1}^T$. Formally, given a data sample from the real distribution $x_0\sim q(x)$, the forward process sequentially adds Gaussian noise over $T$ timesteps:
\begin{align}
    q(x_{1:T}|x_0)=\prod_{t=1}^T q(x_t|x_{t-1});\qquad q(x_t|x_{t-1})=\mathcal{N}(x_t;\sqrt{1-\beta_t}x_{t-1},\beta_t I) \label{eq1}
\end{align}
With the Markov chain process defined in this way, we can sample $x_t$ from $x_0$, denoted as follows:
\begin{align}
    q(x_t|x_0) = \sqrt{\bar{\alpha}_t}x_0 + \epsilon\sqrt{1-\bar{\alpha}_t},\quad \epsilon\sim\mathcal{N}(0,I) \label{eq2}
\end{align}
where $\alpha_t = 1-\beta_t$ and $\bar{\alpha}_t=\prod_{i=1}^t \alpha_t$.

Reversely, the \textit{reverse process}, also called \textit{sample process} is denoising to the raw data. \citet{ho2020denoising} parameterize it as a function $\epsilon(x_t, t)$ which predicts the noise component at timestep $t$. Then the sampling process can be modeled as:
\begin{align}
    p_{\theta}(x_{t-1}|x_t)=\mathcal{N}(x_{t};\mu_{\theta}(x_t),\Sigma_\theta(x_t) I) \label{eq3}
\end{align}
where $\mu_{\theta}=\frac{1}{\sqrt{\alpha_t}}(x_t-\frac{\beta_t}{\sqrt{1-\bar{\alpha}_t}}\epsilon_{\theta}(x_t,t))$ and $\Sigma_{\theta}$ is fixed to a constant.

To sample with semantic or label guidance, \citet{dhariwal2021diffusion} proposed to add an additional classifier to help the posterior sampling $p(x_{t-1}|x_t,y)$ with the target $y$. Inspired by the powerful generation and controllable ability, we use the diffusion model as the backbone network to sample the triplane representation.

\subsection{3D Medical Inverse Problem Solving}
\label{sec2-3}
Considering a forward model for an imaging system, the image inverse problem means estimates the unknown image $x$ given limited measurement $y$, which is challenging due to its ill-posed nature. A general forward imaging model is as follows: 
\begin{align}
\label{eq:mii}
    y&=\mathcal{A}(x)+n\\
    \hat{x}&=D_{\theta}(y) \label{eq:mii2}
\end{align}
where $\mathcal{A}$ denotes the degradation function. Specifically, $\mathcal{A}$ means partial sampling in the sinogram of SV-CT and LA-CT tasks, Poisson disk sampling in $k$-space (i.e. Fourier space) for CS-MRI task, and resolution downsampling along with the spatial $z$-axis (along slice dimension) for ZSR-MRI task. $n$ denotes the measurement noise in the imaging system. $D_{\theta}$ is the parameterized model used to solve the inverse problem and recover the original image $x$ from the degraded measurements $y$.

The two most previous works, DiffuseMBIR~\cite{chung2023solving} and TPDM~\cite{lee2023improving}, both utilize the diffusion model as the generation backbone. Among them, DiffuseMBIR employs the model-based iterative reconstruction (MBIR) method~\cite{katsura2012model} and diffusion model to solve zero-shot medical inverse problem on 3D with diffusion model, which utilizes Total Variation (TV) regularization to constraint the smoothness between adjacent slices. However, simple $z$-axis constraints cannot fully capture the complex 3D distribution, resulting in significant volumetric inconsistencies.
The volume-based model TPDM~\cite{lee2023improving} makes further improvements by modeling the 3D data distribution as a product of two perpendicular 2D plane distributions. And TPDM performs posterior-based sampling \cite{chung2023diffusion} alternatively in perpendicular direction to solve the inverse problem. Despite the significant performance, TPDM still suffers from poor details, artifacts and low efficiency. In our paper, our Blaze3DM also employs the diffusion model as the backbone to generate the triplane embeddings. We propose the guided-based sampling method, inspired by \cite{chung2023diffusion,dhariwal2021diffusion}, which ensures both efficiency and generation quality.

\section{Methodology}

\begin{figure}[t!]
    \centering
    \includegraphics[width=1\linewidth]{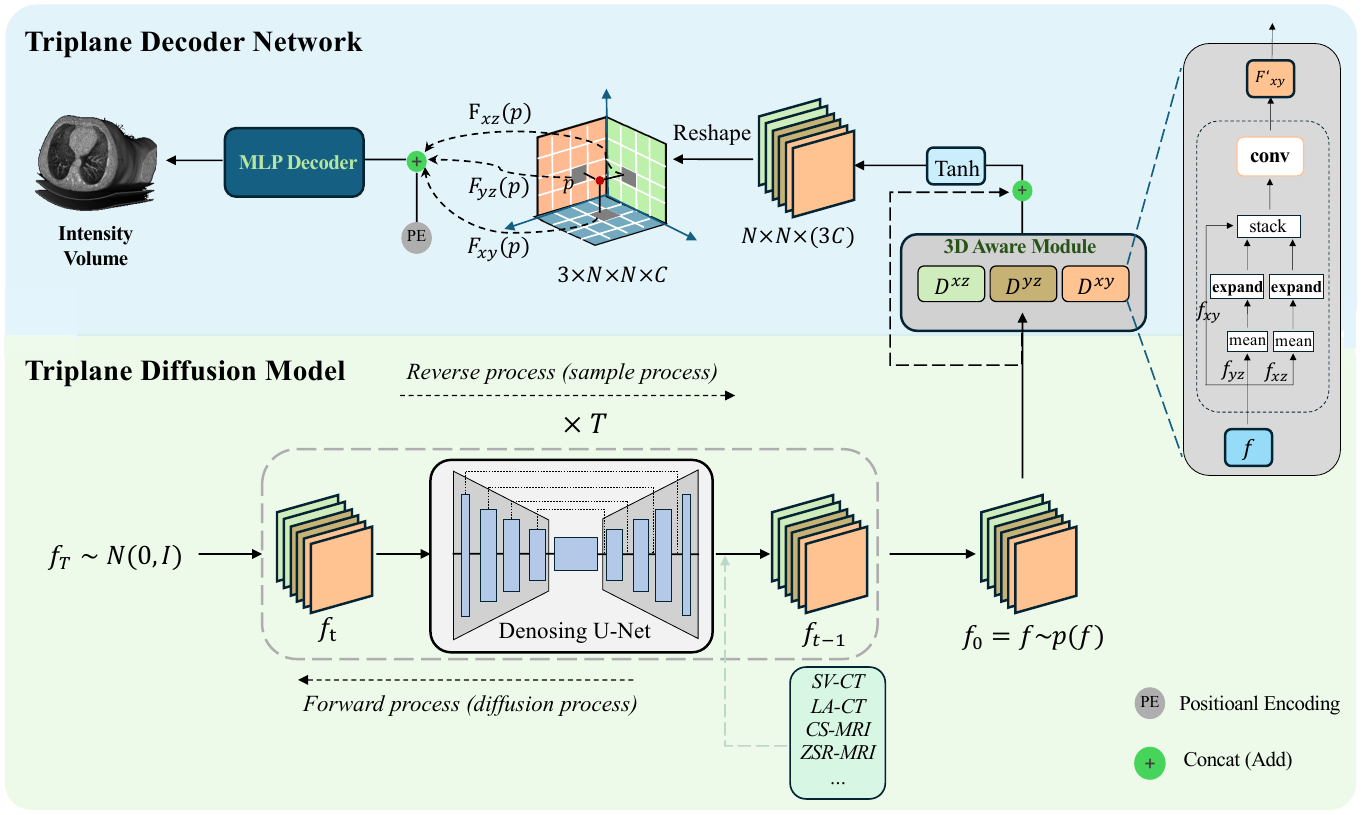}
    \caption{The diagram of Blaze3DM. (a) The top shows the Triplane Decoder Network, which decodes the triplane representation $f$ to the volume intensity. The decoder includes a 3D aware module and a lightweight MLP decoder. (b) The bottom shows the Triplane Diffusion Model, which utilizes the diffusion model to generate triplane embeddings under unconditional/conditional settings.} 
    \label{fig:architecture}

\end{figure}

Let $I\in \mathbb{R}^{H\times W\times D}$ denote the intensity of the medical volume $\mathcal{V}$, where $H, W,D$ are the volume size along $X,Y,Z$ axis. In Section \ref{sec3-1}, we train the triplane representation $f$ of each data and shared decoder network $D_{\phi}(\cdot)$ to reconstruct the medical volume. In Section \ref{sec3-2}, we train the guidance-free diffusion model as the generative backbone for triplane representation, and perform guided sampling to solve various inverse problems. The diagram of Blaze3DM is illustrated in Fig.\ref{fig:architecture}.

\subsection{Triplane Modeling for Medical Volume}

\label{sec3-1}
Triplane representation $f\in \mathbb{R}^{N\times N\times 3C}$, with $N$ as the resolution size and $C$ as the channel count, includes three axis-aligned orthogonal feature planes $f_{xy},f_{yz},f_{xz} \in \mathbb{R}^{N\times N\times C}$, corresponding to $XY$, ${YZ}$ and ${XZ}$ plane, respectively. For each position coordinates $p\in\mathbb[-1,1]^3$ in the normalized space of volume, the summation of projected feature $f_{xy}(p)$, $f_{yz}(p)$,$f_{xz}(p)$ composes the complete latent feature. Triplane not only provides compact representation with efficient information extraction capability but also enables 2D models to learn its distribution.

Each triplane representation can be fed into the shared decoder to render the intensity field (ie. map to medical volume). The decoder is mainly composed of  3D Aware Module and MLP Decoder, as shown in Fig.\ref{fig:architecture}.
\paragraph{3D Aware Module.}
Inspired by Sin3DM~\cite{wu2023sin3dm}, to further enhance expressiveness and maintain volume consistency, we introduce the 3D aware module with residual connection to extract mutual correlation among three orthogonal planes and generate enhanced triplane $F\in \mathbb{R}^{N\times N \times 3C}$, composed by $F_{xy},F_{yz},F_{xz}\in R^{N\times N \times C}$.
\begin{align}
    F_{xy}&=Tanh(D_{\phi}^{xy}(f)+f_{xy})=Tanh(F^{'} _{xy}+f_{xy})\\
    F_{yz}&=Tanh(D_{\phi}^{yz}(f)+f_{yz})=Tanh(F^{'} _{yz}+f_{yz})\\
    F_{xz}&=Tanh(D_{\phi}^{xz}(f)+f_{xz})=Tanh(F^{'} _{xz}+f_{xz})
\end{align}
where $(D_{\phi}^{xy},D_{\phi}^{yz}, D_{\phi}^{xz})$ denote the 3D aware module based on convolution operations. The activation function $Tanh(x)=\frac{e^z-e^{-z}}{e^z+e^{-z}}$ normalizes the range to $[-1,1]$.

Following that, we project the coordinate $p$ into three planes respectively, thus retrieving the corresponding feature vector $F(p) =[F_{xy}(p),F_{yz}(p),F_{xz}(p)]$ via bilinear interpolation. 
\paragraph{MLP Decoder.} Then, we conduct a lightweight MLP network $D_{\phi}^{mlp}$ to predict the volume intensity $\bar{I}(p)$ based on triplane feature $F(p)$ in position $p$. Implicit neural field methods are often limited in expressing fine details due to spectral bias\cite{rahaman2019spectral}. To address this issue, we propose two innovative approaches. Firstly, we integrate Fourier positional encoding and triplane feature of position $p$, enhancing the MLP's capacity to model fine-grained details by leveraging multi-scale frequency information. Secondly, we substitute the traditional activation functions in the MLP with sin functions, which can better capture complex variations and periodic patterns in the image~\cite{molaei2023implicit}.
\begin{align}
    \bar{I}(p)=D_{\phi}^{mlp}([F(p),PosEnc(p)])=D_{\phi}^{mlp}([F_{xy}(p),F_{yz}(p),F_{xz}(p),PosEnc(p)])
\end{align}

\paragraph{Triplane Fitting.}
In the training stage, the triplane representation $f$ of each data and shared decoder network $D_{\phi}$ (including the 3D aware module and MLP network) are both trainable parameters. We optimize them together as autodecoder to reconstruct medical volume by minimizing the MSE (mean square error) loss between the predicted intensity $\bar{I}(p)$ and ground truth $I(p)$:
\begin{align}
    f^*,\phi^* =\mathop{argmin}\limits_{f,\phi} \mathbb{E}_{p\in V }\|I(p)-\bar{I}_{f,\theta}(p)\|^2 \triangleq \mathcal{L}_{rec}
    \label{eq:decloss}   
\end{align}

\paragraph{Loss Regularization.}
Inspired by \cite{shue20233d}, we incorporate three regularization terms to train the decoder network: TV regularization is to enhance information smoothness in the triplane space, L2 regularization is to discourage outlying values of triplane features, and Explicit Intensity Regularization (EIR) to ensure local smoothness in the image space.
\begin{align}
    &\mathcal{L}_{TV}=\mathbb{E}_{p\in V }(\|TV(f_{xy}(p))\|+\|TV(f_{yz}(p))\|+\|TV(f_{xz}(p))\|)\\
    &\mathcal{L}_{L2}=\mathbb{E}_{p\in V }(\|f_{xy}(p)\|^2+\|f_{yz}(p)\|^2+\|f_{xz}(p)\|^2)\\
    &\mathcal{L}_{EIR}=\mathbb{E}_{p\in V }(\|\bar{I}(p)-\bar{I}(p+w)\|), w \in U(p)
\end{align}
where $U(p)$ denotes the spatial neighborhood of position $p$.
Overall, the training objective function of triplane fitting with the above regularization terms is as follows:
\begin{align}
\mathcal{L}_{Dec}=\mathcal{L}_{rec}+\lambda_1\mathcal{L}_{TV}+\lambda_2\mathcal{L}_{L2}+\lambda_3\mathcal{L}_{EIR}
\label{eq:triloss}
\end{align}
where $\lambda_1,\lambda_2,\lambda_3$ are the weights to balance these three regularization terms.

\subsection{Triplane Diffusion Model}

\label{sec3-2}
Once trained the triplane embeddings $\{f^*\}$ and decoder$D_{\phi^*}$, we establish the triplane diffusion model to capture the complex distribution of triplane representation. The diffusion model is formulated in Section \ref{sec2-1}, in which the input $x$ is changed to triplane representation $f^*$. The representation can be unconditionally sampled as follows:
 \begin{align}
    p_{\theta}(f_{t-1}|f_t)=\mathcal{N}(f_{t-1};\mu_{\theta}(f_t),\Sigma_\theta(f_t)I) 
 \end{align}
 \begin{wrapfigure}[14]{r}{0.55\linewidth}
 \vspace{-10mm}
 \begin{minipage}{\dimexpr\linewidth-2\fboxsep-2\fboxrule}

\begin{algorithm}[H]

\caption{Guidance-based Sampling}
  \label{alg:algorithm1}
  \begin{algorithmic}[1]
    \REQUIRE Pretrained decoder $D_{\phi^*}$, degradation transformation $\mathcal{A}$, measurement $y$, gradient scale $\lambda $, guidance slice count $\gamma$.
    \STATE $f_T\sim \mathcal{N}(0,\sigma_1^2 I)$
    \FOR{$t=T\;\text{\textbf{to}}\; 1$}
    \STATE $\mu_{t-1} \leftarrow \mu_{\theta}(f_t);\; \Sigma_{t-1} \leftarrow \Sigma_{\theta}(f_t)$
    \STATE $f_{t-1} \leftarrow$ sample from $\mathcal{N}(\mu_{t-1},\Sigma_{t-1}I)$
    \STATE $V_\gamma \leftarrow$ random select $\gamma$ slice regions
    \STATE $\begin{aligned}
        &{f}_{t-1} \leftarrow f_{t-1} + \\
        &\quad \lambda\Sigma\nabla_{f_{t-1}} \mathbb{E}_{p\in V_\gamma }\left\|\mathcal{A}(D_{\phi^*}(f_{t-1}(p)))-y(p) \right\|^2
      \end{aligned}$
    \ENDFOR
    \STATE ${x} = D_{\phi^*}(f_0)$
    \RETURN ${x}$
  \end{algorithmic}
\end{algorithm}
    \end{minipage}
\end{wrapfigure}
where $Z$ is a normalizing constant. 
\citet{dhariwal2021diffusion} has proven that new distribution under the guidance incorporating can be approximated by a Gaussian distribution with shifted mean:
\begin{align}
    p_{\theta}(f_{t-1}|f_t)p_{\theta}(\mathcal{A},y|f_{t-1})\sim\mathcal{N}(\mu+\Sigma g,\Sigma)
    \label{eq:condshift}
\end{align}
where $\mu = \mu_\theta(f_t), \Sigma=\Sigma_\theta(f_t), g=\nabla_{f_{t-1}}\log p_{\theta,\phi} (\mathcal{A},y|f_{t-1})$. 
Here, the condition in the medical inverse problem forces us to satisfy $\mathcal{A}(D_{\phi^*}(f_0))=y \Leftrightarrow \|\mathcal{A}(D_{\phi^*}(f_0))-y\|^2=0$. We can model the $p_{\theta,\phi} (\mathcal{A},y|f_{0})$ with Gaussian prior:
\begin{align}
    &p_{\theta,\phi} (\mathcal{A},y|f_{t-1})=\frac{1}{\sqrt{2\pi}^n\Sigma^2}exp[-\frac{\|\mathcal{A}(D_{\phi^*}(f_{t-1}))-y\|^2}{2\sigma^2}]\\
\Rightarrow &\nabla_{f_{t-1}}\log p_{\theta,\phi} (\mathcal{A},y|f_{t-1})\approx -\frac{1}{\sigma^2}\nabla_{f_{t-1}}\|\mathcal{A}(D_{\phi^*}(f_{t-1}))-y\|^2
\label{eq:shift}
\end{align}
\paragraph{Guidance-based Sampling.}

As expressed in Section \ref{sec2-3}, the medical inverse problem requires restoring the clean volume given the degradation transformation $\mathcal{A}$ and the measurement $y$. Therefore, our Blaze3DM solves it with this posterior distribution:
\begin{align}
    p_{\theta}(&f_{t-1}|f_t;\mathcal{A},y)\nonumber\\
    &=Zp_{\theta}(f_{t-1}|f_t)p(\mathcal{A},y|f_{t-1})
\end{align}
Combining Eq.\eqref{eq:condshift} and Eq.\eqref{eq:shift}, the conditional sample in the timestep $t$ can be expressed as:

\begin{align}
    f_{t-1}\leftarrow f_{t-1}+\lambda\Sigma \nabla_{f_{t-1}}\|\mathcal{A}(D_{\phi^*}(f_{t-1}))-y\|^2
\end{align}
where $\lambda$ is the guidance scale. 
To save computational costs, we practically randomly select $\gamma$ slice regions in volume space (XY slice for SV-CT, LA-CT, CS-MRI, and XZ/YZ slice for ZSR-MRI) to perform efficient guidance. The pseudo-algorithm of our guidance-based sampling is depicted in Algorithm.\ref{alg:algorithm1}.

\paragraph{Diffusion Training.} Following DDPM~\cite{ho2020denoising}, we train our diffusion model with the simplified objective function:
\begin{align}
    \theta^* = \mathop{\arg\min}\limits_{\theta} \mathcal{L}_{ddpm}=\mathbb{E}_{t,z_0,\epsilon}[&\|\epsilon-\epsilon_{\theta}(\sqrt{\bar{\alpha_t}}z_0+(1-\alpha_t)\epsilon,t)\|^2]
    \label{eq:dmloss}
\end{align}
The triplane representation $f$ consists of three parts $f_{xy}$,$f_{yz}$ and $f_{xz}$, each corresponding to different planes and exhibiting distinct distribution patterns. Consequently, optimizing them at the same scale can lead to gradient explosion problems~\cite{wu2023sin3dm}. To mitigate this issue, we decompose the diffusion training loss into three separate components, as described in Eq\eqref{eq:dmloss}.
\begin{align}
\mathcal{L}_{Diff}=\mathbb{E}_{t,z_0,\epsilon}[&\|\epsilon_{xy}-\epsilon_{\theta}(\sqrt{\bar{\alpha_t}}z_0+(1-\alpha_t)\epsilon,t)_{xy}\|^2\\
&+\|\epsilon_{yz}-\epsilon_{\theta}(\sqrt{\bar{\alpha_t}}z_0+(1-\alpha_t)\epsilon,t)_{yz}\|^2\\
&+\|\epsilon_{xz}-\epsilon_{\theta}(\sqrt{\bar{\alpha_t}}z_0+(1-\alpha_t)\epsilon,t)_{xz}\|^2]
\end{align}

\section{Experiments}

We evaluate our model performance and efficiency on four classic medical inverse problem tasks: sparse-view CT (SV-CT), limited-angle CT (LA-CT), compressed-sensing MRI (CS-MRI) and MRI isotropic super-resolution (ZSR-MRI). We follow the same degradation procedure demonstrated in previous works~\cite{chung2023solving,lee2023improving}.
\begin{figure}[th!]
    \centering
    \includegraphics[width=0.85\linewidth]{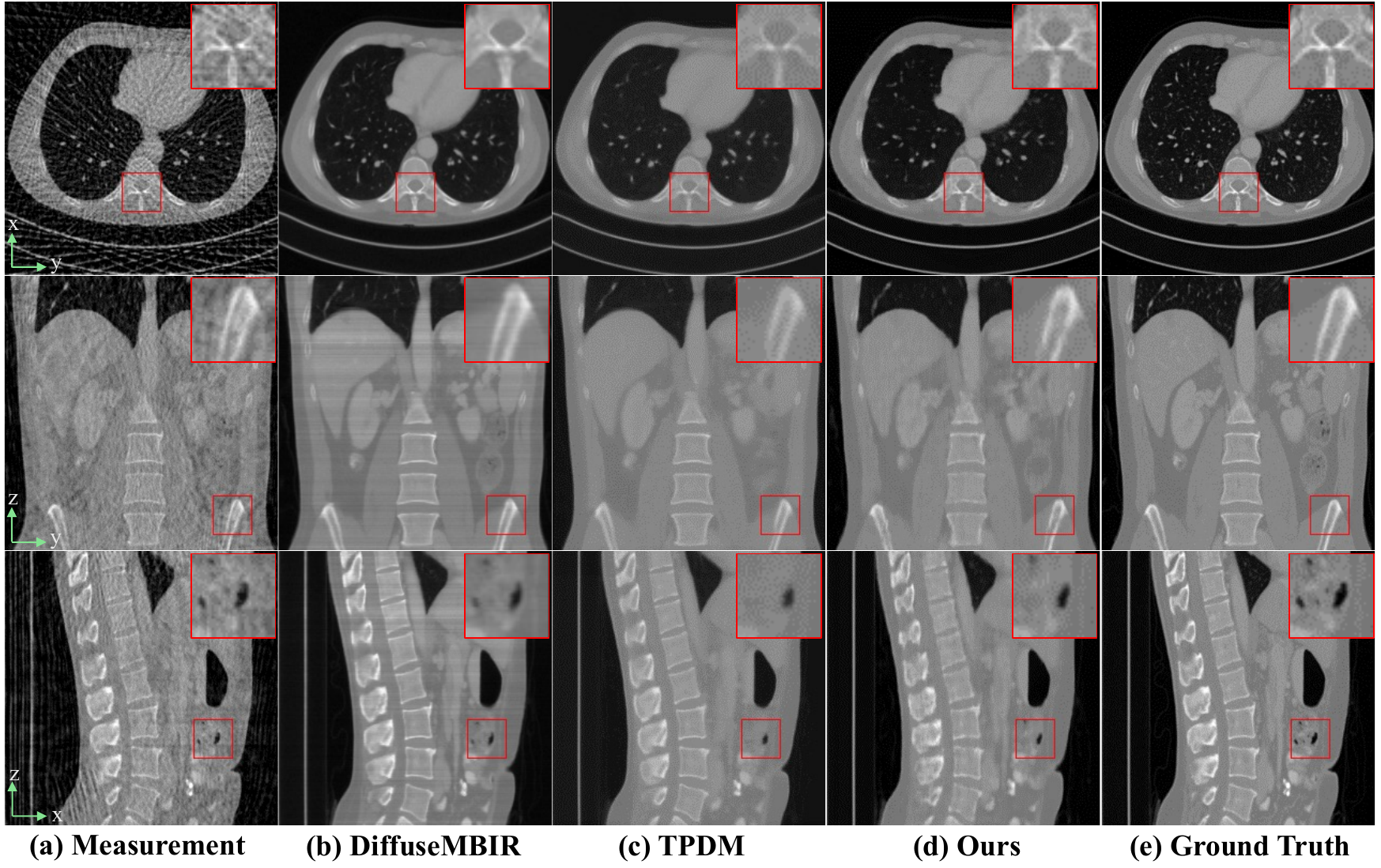}
    \caption{36-view SV-CT reconstruction results of the test volume of AAPM CT dataset.  (\textbf{First row}: axial plane; \textbf{ Second row}: coronal plane; \textbf{Third row}: sagittal plane)}
    \label{fig:svCT}

\end{figure}

\paragraph{CT Dataset.} Following previous work~\cite{chung2023solving,lee2023improving}, we used the public CT dataset provided in AAPM 2016 CT low-dose grand challenge~\cite{mccollough2017low}. The dataset consists of a total of 10 volumes of contrast-enhanced abdominal CT. We divided the train-test dataset the same as TPDM~\cite{lee2023improving}, using 9 volumes to train and one volume to test. We use the original 3D medical image without cropping or resizing. The degradation uses uniform 36 views for SV-CT and limited 0-90 degrees for LA-CT, respectively.

\paragraph{MRI Dataset.} For MRI tasks, we use the public IXI dataset\footnote{\url{http://brain-development.org/ixi-dataset/}}, which contains 3D multi-modal human brain scans, where each MRI has three modalities (T1, T2 and PD-weighted). Here, we only use the T1-weighted modality data for training and testing,  and resize volumes to $256\times 256\times 256$. For CS-MRI, we perform reconstruction on retrospective ×8 acceleration Poisson sub-sampled CS-MRI volumes. For ZSR-MRI, the measurement degradation is $4×$ anisotropy along $z$-axis. 

\paragraph{Evaluation Metrics.} We take two image quality metrics for evaluation. The Peak Signal Noise Ratio (PSNR) measures pixel-wise value similarity between images, while SSIM~\cite{wang2004image} reflects the human visual perception of local structural similarity between images.

\subsection{Model Implementation Details} 

\label{sec:exp_detail}
\textbf{Triplane Fitting}. The triplane size is set to $(128,128,32)$ by default. The weights in Eq.\eqref{eq:triloss} are typically set to $\lambda_1=0.01,\lambda_2=0.001,\lambda_3=1$. Adam optimizer ($\beta_1 =0.9, \beta_2 = 0.999$) was taken with a learning rate of $1e-3$ and a batch size of 1. Random resize augmentation was used to enhance the model’s scalability.

\textbf{Diffusion Training}. We follow the parameter settings of guided diffusion described in its official usage guidance \cite{dhariwal2021diffusion}. To accelerate the convergence of the diffusion model, we use the NFD~\cite{shue20233d} pretrained checkpoint on the ShapeNet dataset~\cite{chang2015shapenet} as initialization. The learning rate of the diffusion model is set to $1e-5$ and the batch size to 1.

\textbf{Posterior Sampling}. The diffusion sampling step is set to 1000. The random slice count $\gamma$ is set to 16, the maximum acceptable value of a 24G GPU card. The scale weight is initialized as $\lambda=6$.

\textbf{Computation Platform.} The code is based on PyTorch framework. We perform all the training and inference experiments on a single NVIDIA RTX 4090 GPU. 

\begin{figure}[h]
    \centering
    \includegraphics[width=0.85\linewidth]{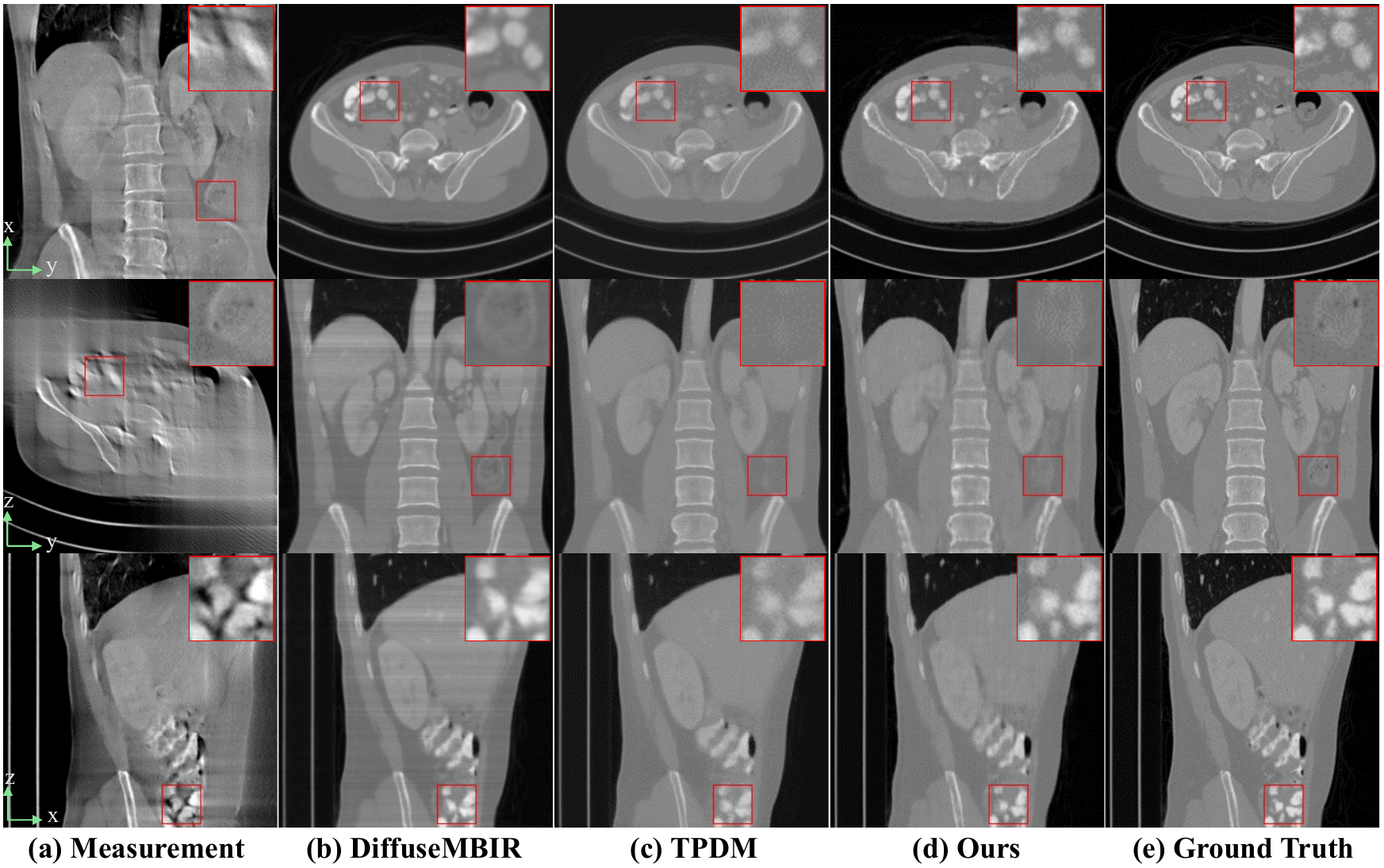}
    \caption{$90\degree$ LA-CT reconstruction results of the test volume of AAPM CT dataset. (\textbf{First row}: axial plane; \textbf{ Second row}: coronal plane; \textbf{Third row}: sagittal plane)}
    \label{fig:laCT}

\end{figure}

\subsection{Main Results}

This section presents our main results compared with DiffuseMBIR~\cite{chung2023solving} and TPDM~\cite{lee2023improving} on the four tasks. Due to space limitations, more CS-MRI, ZSR-MRI and triplane results are shown in Appendix.

\paragraph{Sparse-view CT (SV-CT) and Limited-angle CT (LA-CT)}
The quantitative results of 36-views CT and  $90\degree$ limited angle CT in Tab\ref{tab:CT} show that our Blaze3DM outperforms DiffuseMBIR~\cite{chung2023solving} and TPDM~\cite{lee2023improving} in both PSNR and SSIM. The visualization results of SV-CT and LA-CT are shown in Fig.\ref{fig:svCT} and Fig.\ref{fig:laCT} respectively. DiffuseMBIR tends to produce layer artifacts along slice direction despite TV regularization. TPDM-produced samples appear overly smooth, which may lose structures in weak contrast regions. Compared to them, Blaze3DM maintains good 3D consistency and delivers superior details that closely match the ground truth.

\begin{table}[th]
\small
\caption{Quantitative evaluation (PSNR, SSIM) of SV-CT (36 views) and LA-CT (90 degrees) on the AAPM test volume. $\uparrow$ denotes the higher the better.}
\centering
\setlength{\tabcolsep}{2pt}
\begin{tabular}[width=0.6\linewidth]{lcccccccc}
\toprule
& \multicolumn{4}{c}{SV-CT} & \multicolumn{4}{c}{LA-CT} \\
\cmidrule(lr){2-5}\cmidrule(lr){6-9}

\multirow{2}{*}{{Methods}}& \multirow{2}{*}{PSNR $\uparrow$ } & \multicolumn{3}{c}{SSIM $\uparrow$} & \multirow{2}{*}{PSNR $\uparrow$}  & \multicolumn{3}{c}{SSIM $\uparrow$} \\ 
 \cmidrule(lr){3-5}  \cmidrule(lr){7-9}
&    & Axial         & Coronal        & Sagittal  &   & Axial         & Coronal        & Sagittal        \\ 
\midrule
DiffuseMBIR~\cite{chung2023solving} & 34.78
& 0.857& 0.856& 0.861 & 34.64& 0.854& 0.852& 0.849
\\ 
\midrule 
TPDM~\cite{lee2023improving}&  {38.25}&  {0.947}&  {0.951}&  {0.949}&  {38.02}
&  {0.947} &  {0.949} &  {0.948}
\\
 \midrule
 Blaze3DM (Ours)    & \textbf{38.39}& \textbf{0.951}& \textbf{0.951}& \textbf{0.951} & \textbf{38.13}
&\textbf{ 0.950 }& \textbf{0.950} & \textbf{0.951} 
\\ 
\bottomrule
\end{tabular}
\label{tab:CT}

\end{table}

\begin{table}[h!]

\small
\caption{Quantitative evaluation (PSNR, SSIM) of CS-MRI and ZSR-MRI. N/W: Not Working. $\uparrow$ denotes that the higher, the better.}
\centering
\setlength{\tabcolsep}{2pt}
\begin{tabular}[width=0.8\linewidth]{lccccccccc}
\toprule
& \multicolumn{4}{c}{CS-MRI} & \multicolumn{4}{c}{ZSR-MRI} \\
\cmidrule(lr){2-5}\cmidrule(lr){6-9}

\multirow{2}{*}{{Methods}}& \multirow{2}{*}{PSNR $\uparrow$ } & \multicolumn{3}{c}{SSIM $\uparrow$} & \multirow{2}{*}{PSNR $\uparrow$}  & \multicolumn{3}{c}{SSIM $\uparrow$} \\ 
 \cmidrule(lr){3-5}  \cmidrule(lr){7-9}
&    & Sagittal         & Axial        & Coronal  &   & Sagittal        & Axial        & Coronal       \\ 
\midrule
DiffuseMBIR~\cite{chung2023solving} & 37.21
&0.939 & 0.933& 0.934 &  \multicolumn{4}{c}{N/W} 
\\ 
\midrule 
TPDM~\cite{lee2023improving}&  {39.18}
& {0.953} &  {0.951}&  {0.950} & 38.42 & 0.948& 0.946&0.946  
\\
 \midrule
 Blaze3DM (Ours)  & \textbf{39.35}
&\textbf{0.957} & \textbf{0.952}& \textbf{0.951} &\textbf{38.86} &\textbf{0.951} & \textbf{0.949}&\textbf{0.949}  
\\ 
\bottomrule
\end{tabular}
\label{tab:MRI}

\end{table}
\paragraph{Compress sensing MRI (CS-MRI) and MRI isotropic super-resolution (ZSR-MRI)}
The quantitative results are present in Tab.\ref{tab:MRI}. The qualitative visualization results are shown in the Appendix. Consistent with the results on SV-CT and LA-CT, Blaze3DM 
\begin{wraptable}[8]{r}{0.6\textwidth}
\centering

\caption{Comparison of inference time, memory and FLOPs with volume size of $256\times 256\times 256$.}
\centering
\scalebox{0.95}{
\begin{tabular}{lccc} 
\toprule
\textbf{Methods} &Time (h)& Memory (G) & FLOPs (G)\\ 
\midrule 
DiffuseMBIR~\cite{chung2023solving}&18±2&  16.2±0.2 & 663.3\\
TPDM~\cite{lee2023improving}&32±3&  21.5±0.3 & 661.0\\ 
Blaze3DM&\textbf{0.8±0.2}& 22.1±0.3 & \textbf{349.7}\\
\bottomrule
\end{tabular}}
\label{tab:efficiency}
\end{wraptable}
also demonstrated better performance than previous works on CS-MRI and ZSR-MRI tasks, accurately restoring details and edges while keeping 3D consistency.

\paragraph{Efficiency Comparison}
Importantly, our triplane neural field diffusion model greatly speeds up the inference time compared to previous methods, shown in Tab.\ref{tab:efficiency}. For unified comparison, the volume size is set to $256\times256\times256$, and the inference is performed on the maximum batch size supported by a single RTX 4090 GPU with $24G$ memory. 
The Blaze3DM achieves almost $22\times$ faster speed than DiffuseMBIR
and $40\times$ faster than TPDM, while having about half  FLOPs (floating point operations). It should be noted that our method is scalable to generate arbitrary size volume, only changing coordinate sets that input to decoder without influencing diffusion sampling. While DiffuseMBIR and TPDM work on fixed-size images, need to be retrained to handle different data sizes, and sampling complexity increases almost linearly with the volume size $\mathcal{O}(H \times W \times C)$.

\subsection{Ablation studies}

\label{sec:appabl}
\begin{table}[t]
\addtolength{\tabcolsep}{-2pt}
    \centering
    \tiny
    \caption{Ablation studies on following important components of Blaze3DM: (a) Channel size of triplane; (b) Resolution size of triplane; (c) Using 3D aware module (3DAM) or not; (d) Regularization terms in the loss function (Eq.\eqref{eq:triloss}).}

    \label{tab:ablation}
    \begin{subtable}[t]{0.2\textwidth}
        \centering
        \begin{tabular}{l|cc}
        \hline
            Cha. & PSNR $\uparrow$ & SSIM $\uparrow$ \\
            \hline
            16 & 37.00 & 0.929 \\
            32 & 38.39 & 0.951 \\
            64 & \textbf{38.93} & \textbf{0.957}\\
            \hline
        \end{tabular}
        \caption{Channel size.}
        \label{tab:abl_cha}
    \end{subtable}\hfill
    \begin{subtable}[t]{0.2\textwidth}
        \centering
        \begin{tabular}{l|cc}
        \hline
            Res. & PSNR $\uparrow$ & SSIM $\uparrow$ \\
            \hline
            64 & 37.77 & 0.939 \\
            128 & 38.39 & 0.951 \\
            256 & \textbf{38.48} & \textbf{0.954} \\
            \hline
        \end{tabular}
        \caption{Resolution size.}
        \label{tab:abl_res}
    \end{subtable}\hfill
    \begin{subtable}[t]{0.2\textwidth}
        \centering

        \begin{tabular}{l|cc}
        \hline
            Case & PSNR $\uparrow$ & SSIM $\uparrow$ \\
            \hline    
            w/o 3DAM & 37.06& 0.933\\
            w/  \;3DAM & \textbf{38.39} & \textbf{0.951} \\
                    \hline
        \end{tabular}

        \caption{3D aware module.}
        \label{tab:abl_3dam}
    \end{subtable}\hfill
    \begin{subtable}[t]{0.4\textwidth}
        \centering

        \begin{tabular}{l|cc}
        \hline
            Case & PSNR$\uparrow$ &SSIM$\uparrow$ \\
            \hline
            None & 35.37 & 0.851 \\
            $+\mathcal{L}_{TV}$ & 38.27 & 0.936 \\
            $+\mathcal{L}_{TV}+\mathcal{L}_{L2}$ & 38.32 & 0.944 \\
            $+\mathcal{L}_{TV}+\mathcal{L}_{L2}+\mathcal{L}_{EIR}$ & \textbf{38.39} & \textbf{0.951}\\
            \hline
        \end{tabular}
        \caption{Regularization terms.}
        \label{tab:abl_reg}
    \end{subtable}

\end{table}

\begin{figure}[th!]
    \centering

    \includegraphics[width=0.85\linewidth]{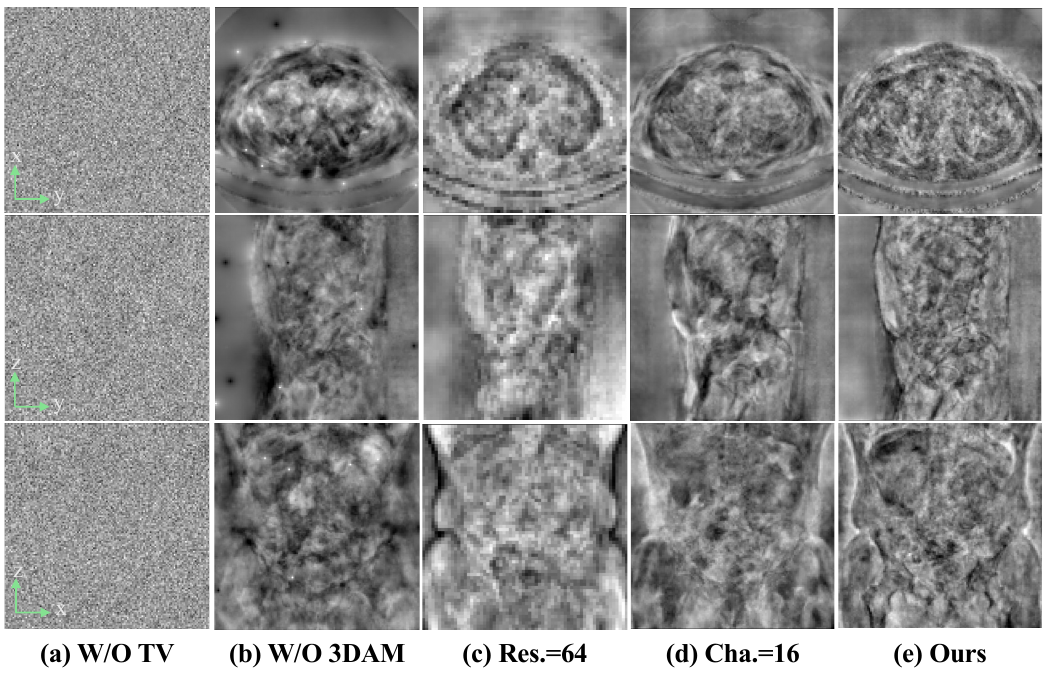}
    \caption{Ablation studies of triplane fitting on AAPM-CT dataset, take triplane of L333 as an example to show. (\textbf{First row}: axial plane; \textbf{Second row}: coronal plane; \textbf{Third row}: sagittal plane)}
    \label{fig:abl}

\end{figure}

In this section, we perform the ablation experiments of triplane fitting on the AAPM CT dataset by ablating the following important components. The evaluation results are shown in Tab.\ref{tab:ablation} and visual comparison is provided in Fig.\ref{fig:abl}.                   

\paragraph{Triplane resolution and channel.}
The resolution and channel size of triplane have significant impact on its expressive ability. Tab.\ref{tab:abl_cha} varies the channels in $\{16,32,64\}$, and shows the performance improves when the channel count gets more. Also, the performance gets better when the resolution increases as $\{64,128,256\}$. Both these results indicate that increased representation size helps express complex information.

\paragraph{3D Aware Module.} As shown in Tab.\ref{tab:abl_3dam}, we compared Blaze3DM with and without the 3D aware module (3DAM). The results in metrics and visualization in Fig.\ref{fig:abl} demonstrate that 3D aware module makes model more expressive and easy to fit, otherwise resulting in dark or bright spots in certain positions.

\paragraph{Regularization terms.}
We use three regularization terms in Blaze3DM as described in Eq.\eqref{eq:triloss}. To evaluate their effectiveness, we sequentially added $\mathcal{L}_{TV}$, $\mathcal{L}_{L2}$, and $\mathcal{L}_{EIR}$ to the base loss function $\mathcal{L}_{rec}$. The results in Tab.\ref{tab:abl_reg} show progressively improved PSNR and SSIM values. Notably, TV regularization is important to keep smoothness in triplane and maintain similar shape with corresponding volume, otherwise the triplane shows in a noisy pattern.

\section{Discussion and Conclusion}
\label{sec:dis}

\paragraph{Contributions and Limitations}
Our proposed Blaze3DM provides a promising direction for fully modeling massive 3D medical volume with triplane representation, which can address the low efficiency, artifact and over-smooth problems in existing methods. In 3D medical inverse problem solving, Blaze3DM is over 22$\times$ faster in inference speed compared to previous methods (Tab. \ref{tab:efficiency}). However, our method requires a known degradation formulation as guidance, which has limitations in blind degradation commonly seen in real-world scenarios.

\paragraph{Conclusion}
In this paper, we propose Blaze3DM, an efficient 3D medical volume generative model for inverse problem solving. The core of our method lies in utilizing a compact triplane neural field to construct a novel representation of 3D medical volume data. Additionally, we employ a powerful diffusion model as the generative backbone to synthesize high-fidelity triplane representations. Furthermore, we introduce a guidance-based sampling method to solve 3D medical inverse problems in zero-shot manner. Extensive experiments demonstrated that Blaze3DM achieves 22\textasciitilde40$\times$ speedup in inference time compared to existing methods, meanwhile resolving issues of volumetric inconsistency and various artifacts present in previous approaches to a certain extent.

\newpage
\medskip
{
\small
\bibliographystyle{unsrtnat}
\bibliography{egbib}


}

\newpage

\appendix

\section{Appendix / supplemental material}

\subsection{MRI results}
\label{sec:appmri}
\begin{figure}[t]
    \centering
    \includegraphics[width=1\linewidth]{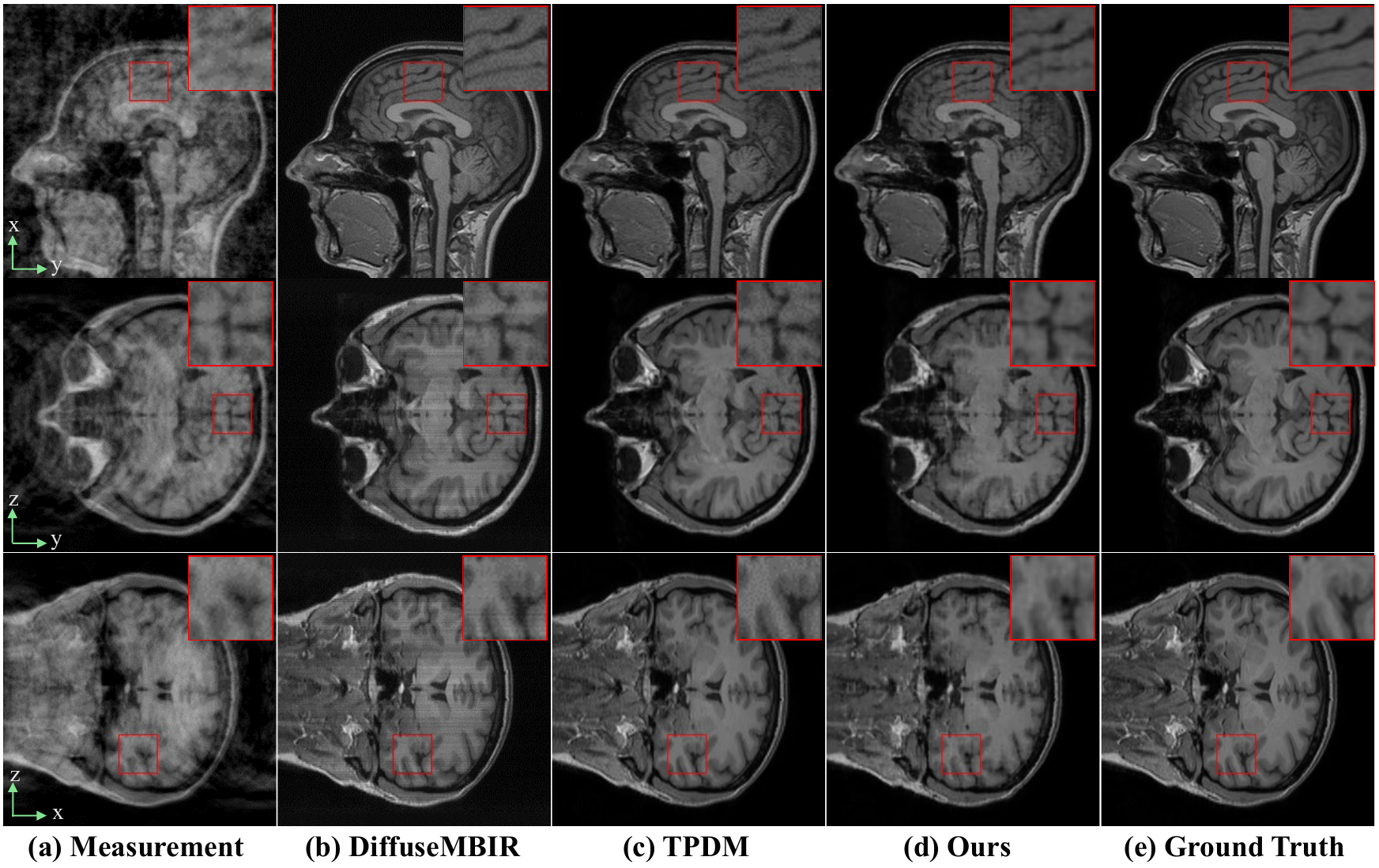}
    \caption{×8 acceleration Poisson sub-sampled CS-MRI reconstruction results of the test volume of IXI dataset (\textbf{First row}: sagittal plane, \textbf{second row}: axial plane, \textbf{third row}: coronal plane).}
    \label{fig:cs-mri}
\end{figure}

\begin{figure}[!h]
    \centering
    \includegraphics[width=1\linewidth]{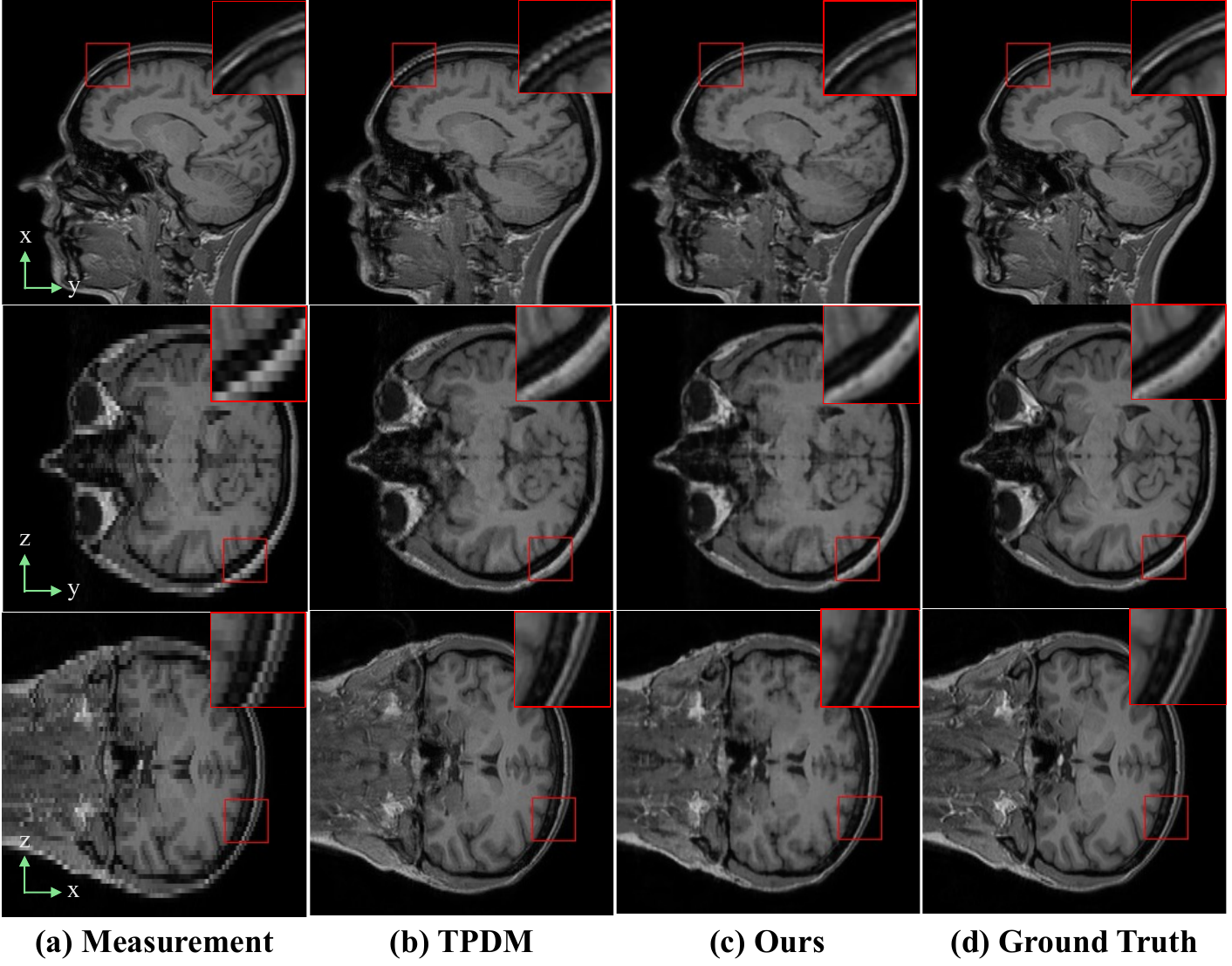}
    \caption{MR-ZSR results (×4 anisotropy) of the test volume of IXI dataset \textbf{(First row:} sagittal plane, \textbf{second row}: axial plane, \textbf{third row:} coronal plane). }
    \label{fig:zsr-mri}
\end{figure}

We present the qualitative visualization results of CS-MRI and ZSR-MRI in Fig.\ref{fig:cs-mri} and Fig.\ref{fig:zsr-mri}.
\paragraph{CS-MRI} We compared our method with DiffuseMBIR and TPDM in the CS-MRI task, and the qualitative results are present in Fig.\ref{fig:cs-mri}. The DiffuseMBIR method, due to insufficient volumetric consistency constraints, exhibits layered artifacts in both the XZ/YZ planes. Both the TPDM and our Blaze3DM methods perform well in volume consistency, while Blaze3DM outperforms significantly in efficiency.

\paragraph{ZSR-MRI} Since the DiffuseMBIR is not applicable to ZSR-MRI task, we compared our method with the TPDM. Both Blaze3DM and TPDM generally exhibit good visualization results in Fig.\ref{fig:zsr-mri}. However, TPDM suffers from severe jagged artifacts in the edge details of the $XY$-plane (as marked by the red boxes), which may caused by inefficient constraints from the auxiliary model in this plane. In contrast, our Blaze3DM method demonstrates excellent detail in every plane, owing to the powerful expressiveness of our triplane neural field model.

\begin{figure}[ht]
    \centering
    \includegraphics[width=1\linewidth]{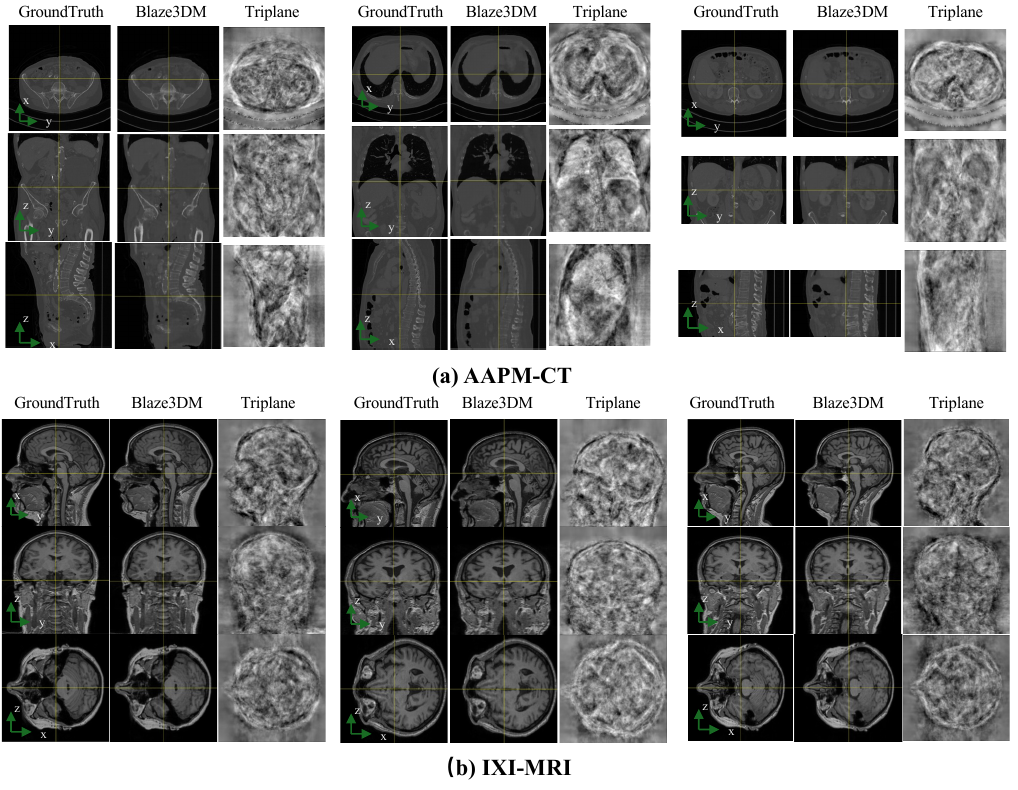}
    \caption{More triplane visualizations of CT and MRI medical image. In each pair, the orthogonal view of 3D volume is shown in left, representative triplane in the middle, and the reconstruction from triplane in right. }
    \label{fig:samples}
\end{figure}

\subsection{More triplane visualization}
\label{sec:appvis}
In this section, we show more triplane visualization of our method. In Fig.\ref{fig:samples}(a), we present three AAPM-CT examples, including the ground truth, our reconstruction and the corresponding triplane representation. In Fig.\ref{fig:samples}(b), another three IXI MRI data examples are shown. 

These results indicate that our method excels in generation quality in details, which comes from that the triplane representation maintains structural consistency with the original 3D image data. 

More importantly, Blaze3DM can be applied to volumes of different size, demonstrating excellent data scalability, a capability that other methods do not possess.

\end{document}